\documentclass{article}

\usepackage{microtype}
\usepackage{graphicx}
\usepackage{subfigure}
\usepackage{booktabs} 
\usepackage{multirow}

\usepackage{hyperref}

\PassOptionsToPackage{hyphens}{url}
\usepackage{amsmath}
\usepackage{amssymb}
\usepackage{mathtools}
\usepackage{amsthm}

\usepackage[preprint]{icml2026}

\usepackage{microtype}
\usepackage{xcolor}

\icmltitlerunning{Auditing Social-Desirability Bias in LLM Annotators}

\begin{document}

\twocolumn[
\icmltitle{Two Wrongs, No Right: Auditing Social-Desirability Bias \\
           in LLM Annotators for Computational Social Science}

\begin{icmlauthorlist}
\icmlauthor{Varun Kotte}{ind}
\end{icmlauthorlist}

\vskip -0.10in
\begin{center}\textit{Independent Researcher}\end{center}
\vskip 0.10in

\icmlaffiliation{ind}{Independent Researcher}

\icmlcorrespondingauthor{Varun Kotte}{kottevarun@gmail.com}

\icmlkeywords{trustworthy AI, LLM annotation, computational social science, alignment, social desirability bias, civic discourse}

\vskip 0.3in
]

\makeatletter
\renewcommand{\Notice@String}{%
  \textit{Preprint. A workshop version of this paper has been submitted to the ICML 2026 Trustworthy AI for Good Workshop.}%
}
\makeatother
\printAffiliationsAndNotice{}

\begin{abstract}
LLM annotators are increasingly used in computational social science (CSS), but it is unclear whether their alignment-shaped errors preserve the empirical conclusions a researcher would report. We audit three open-source 7B instruction-tuned models (Zephyr, Mistral-Instruct, Qwen2.5-Instruct) across six TweetEval tasks under four prompt conditions (72 cells) and find that social-desirability failures do not run in a single direction. Zephyr exhibits \emph{leniency bias}, systematically under-applying harmful labels (offensive language: false benign rate $0.729$, false alarm rate $0.031$). Mistral and Qwen exhibit \emph{overcorrection}, over-applying the same labels (Mistral hate-speech FAR $=0.604$). All three models exhibit \emph{neutrality bias} on abortion stance, underestimating opposition prevalence by $24$ to $40$ percentage points and inflating the neutral label. None of the four prompting interventions we test (neutral, safety framing, depersonalized, chain-of-thought) corrects these failures across models; safety framing can worsen stance distortion. Strikingly, Zephyr's hate-speech prevalence estimate matches the gold rate exactly while its class-conditional errors are large in both directions, an accidental cancellation that misleads aggregate validation. We translate these patterns into a three-part taxonomy with diagnostic FBR/FAR signatures and a lightweight gold-sample validation protocol. The headline for trustworthy CSS: a model that looks calibrated on aggregate metrics can still flip the substantive empirical conclusion a researcher would report.
\end{abstract}

\section{Introduction}
\label{sec:intro}

LLMs are now routinely proposed as low-cost annotators for computational social science (CSS), where labels feed political-text studies, hate-speech monitoring, sentiment tracking, and public-opinion measurement \citep{gilardi2023chatgpt,ziems2024can,tornberg2025large}. Because these labels are increasingly used to inform civic discourse and policy debate, errors in the annotator become errors in the social-scientific record, and ultimately in the picture of public sentiment that institutions and the public use to make decisions. A trustworthy AI account of LLM annotation must therefore look past per-instance accuracy and ask whether automated labels preserve the empirical quantities a researcher would actually report.

That distinction matters because most CSS pipelines use annotation as a \emph{measurement} step: estimating prevalence, comparing groups, tracking temporal shifts, or testing associations between text-derived variables and political, demographic, or behavioral outcomes. In such settings, a model can have superficially acceptable average accuracy while systematically under-producing or over-producing a socially salient label. Recent work shows that aligned LLMs exhibit prompt-sensitive response biases in survey and opinion settings, especially after preference optimization \citep{tjuatja2024llms,santurkar2023whose,bang2024measuring}. We still lack a clear task-level account of how those tendencies appear in the specific annotation tasks CSS researchers deploy.

We focus on \emph{social-desirability response bias} (SDR): the tendency to produce responses that are normatively preferred rather than maximally faithful to the underlying instance \citep{tjuatja2024llms}. SDR need not look like a single bias direction. A model that is reluctant to emit a harmful label may undercount hateful content, while a differently trained model may over-apply the same label to avoid seeming permissive. On politically sensitive tasks, a model may soften strong positions by drifting toward a balanced-looking neutral answer.

\textbf{Contributions.} We study this question through a comparative audit spanning six TweetEval tasks \citep{barbieri2020tweeteval}, three open-source 7B instruction-tuned models, and four prompting conditions, yielding 72 experimental cells. The three models differ along several training axes simultaneously, so our design is best read as a contrast set rather than a controlled ablation. We make three contributions.
\begin{enumerate}\itemsep0pt
\item A multi-task audit of alignment-sensitive annotation behavior across hate speech, offensive language, stance, sentiment, emotion, and irony.
\item Evidence that aggregate evaluation is insufficient for CSS deployment: pairing class-conditional error with downstream prevalence distortion exposes cases where prevalence accuracy is misleadingly correct due to offsetting errors.
\item A descriptive taxonomy of three recurring failure patterns (leniency bias, overcorrection, neutrality bias) with diagnostic criteria and a validation protocol that practitioners can run before scaling an annotation pipeline.
\end{enumerate}

The takeaway for trustworthy AI for societal good: when LLMs label text used to characterize public discourse, opposing failure modes can, on the same task, push corpus-level conclusions in opposite directions, as our hate-speech results illustrate. Reporting the model and the prompt as part of the measurement instrument is therefore a baseline requirement, not an optional disclosure.

\section{Related Work}

\textbf{LLMs as CSS annotators.}
\citet{gilardi2023chatgpt} show that ChatGPT can outperform crowd workers on selected political text annotation tasks, while \citet{ziems2024can} provide a broader evaluation across 25 CSS benchmarks and conclude that LLMs are promising but uneven, particularly on classification tasks. \citet{pangakis2023automated} argue that LLM-based annotation must be validated task by task rather than assumed reliable by default. \citet{he2024annollm} propose explanation-augmented prompting as a quality intervention. A parallel and rapidly growing literature studies LLMs as evaluators or judges and documents structurally similar bias patterns such as position, length, and self-preference bias \citep{gu2024survey}; our taxonomy offers the annotator-side analog of those judgment biases. Our work moves this discussion from per-instance accuracy to task-level error profiles aligned with downstream measurement.

\textbf{Response bias and alignment.}
LLM outputs are shaped not only by pretraining data but also by post-training procedures such as supervised instruction tuning and preference optimization. \citet{tjuatja2024llms} show that LLMs exhibit prompt-sensitive response biases in survey settings and that RLHF-ed models often diverge from human response patterns. \citet{santurkar2023whose} find that language-model opinion profiles reflect systematic ideological and demographic skews, and \citet{bang2024measuring} demonstrate that political bias depends on both content and framing. Recent theoretical work argues that alignment objectives may inherently push aligned models toward particular ideological priors \citep{hagendorff2025inevitability}; our empirical finding is that, on a contested political annotation task, this tendency manifests as drift away from strong stance positions toward a neutral-looking distribution rather than as a uniform ideology-coded answer. A complementary line of work shows that LLM outputs can vary substantially at the format level across both models and prompts in ways that semantic accuracy metrics may not capture \citep{kotte2026promptport}. We connect this literature to annotation deployment by isolating directional error in the labels themselves.

\textbf{Bias in annotation pipelines.}
Annotation is not a neutral channel: human annotator identity and beliefs affect labels in harmful-language settings \citep{sap2022annotators}, and bias enters NLP systems at multiple stages of the pipeline \citep{hovy2021five}. More recent audits of LLM annotators on hate-speech tasks document systematic group-level biases across frontier model families \citep{das2024annotator}; our work is complementary in that we focus on the \emph{direction} of error (under- versus over-application of the harmful label) and its consequences for prevalence estimates rather than on demographic group disparities. Comparative analyses of human and LLM annotation biases on hate-speech tasks find that LLM biases differ structurally from those of human annotators across socio-demographic axes \citep{giorgi2024human}, supporting our claim that aligned-LLM error profiles cannot be assumed to follow the noise distribution of human annotation. A parallel strand on annotator disagreement and label subjectivity argues that treating annotation as a gold-standard consensus process obscures genuine variation in human judgment \citep{plank2022problem,davani2022dealing}. Our contribution is not to claim that LLMs uniquely introduce bias, but to show that aligned LLM annotators produce specific high-impact error \emph{profiles} that are easy to miss if one reports only overall accuracy or macro-F1, and that differ systematically from the distributional noise associated with human annotator disagreement.

\textbf{Measurement validity and construct distortion.}
From a CSS perspective, the most relevant failure is often not misclassification in isolation but distortion of the latent construct a researcher intends to measure. Work in social-scientific measurement has long emphasized that operationalizations should be evaluated by whether they preserve the intended construct and support valid downstream inference, not merely by face-valid individual decisions \citep{jacobs2021measurement}. Recent discussions of LLM-assisted labeling make a similar point: automated coders may appear competent on average while shifting prevalence estimates, treatment effects, or correlational patterns in systematically misleading ways \citep{argyle2023out,pangakis2023automated}. Recent work attempts to address this distortion directly by specializing models to match survey response distributions \citep{cao2025specializing}; we contribute the diagnostic counterpart, showing how to detect such distortion in off-the-shelf annotators before deployment. Our paper operationalizes that concern in a compact audit setting by pairing standard task metrics with class-conditional and prevalence-sensitive measures.

\section{Experimental Setup}

\subsection{Tasks and data}
We evaluate six tasks from TweetEval \citep{barbieri2020tweeteval}: hate speech, offensive language, stance detection on abortion, sentiment, emotion, and irony. These tasks span both high-SDR settings (where one class label is itself socially undesirable to emit, such as \textsc{hate speech}, \textsc{offensive}, or a strong political stance) and lower-SDR settings such as sentiment and emotion. We use the published test splits throughout. Datasets with more than 500 test examples are subsampled to 500 instances using a fixed seed of 42; the abortion stance test set contains 280 instances. Table~\ref{tab:tasks} summarizes the task inventory.

\begin{table}[t]
\caption{Evaluation tasks from TweetEval.}
\label{tab:tasks}
\centering
\small
\begin{tabular}{lcc}
\toprule
Task & Labels & $N$ \\
\midrule
Hate speech & non-hate / hate speech & 500 \\
Offensive language & non-offensive / offensive & 860 \\
Stance (abortion) & none / against / favor & 280 \\
Sentiment & negative / neutral / positive & 500 \\
Emotion & anger / joy / optimism / sadness & 500 \\
Irony & non-irony / irony & 500 \\
\bottomrule
\end{tabular}
\end{table}

\subsection{Models}
We evaluate three openly available 7B instruction-tuned models that differ in base family and post-training recipe: Zephyr-7B \citep{tunstall2023zephyr}, Mistral-7B-Instruct-v0.3 \citep{jiang2023mistral}, and Qwen2.5-7B-Instruct \citep{qwen2025qwen25}. Zephyr is a heavily preference-optimized chat model, Mistral-Instruct combines instruction tuning with subsequent preference optimization, and Qwen includes supervised instruction tuning plus additional alignment stages. We treat the trio as a practical contrast set rather than a clean causal ablation over a single training variable. All inference uses greedy decoding at temperature $0$ for reproducibility, run in float16 on a single NVIDIA A100 80GB GPU.

\subsection{Prompting conditions}
Each example is annotated under four system-message conditions: (i) \textbf{Neutral} (task instruction only); (ii) \textbf{Safety framing} (adds explicit language about safety, fairness, and avoiding harm); (iii) \textbf{Depersonalized} (presents the model as a deterministic labeling function rather than a helpful assistant); (iv) \textbf{Chain-of-thought} (allows brief intermediate reasoning before the final label) \citep{wei2022chain}. These conditions reflect interventions a practitioner could plausibly try without additional model training. Full prompt templates appear in Appendix~\ref{app:prompts}.

\subsection{Label parsing and metrics}
Prompts enumerate the legal label set and require the final answer to contain exactly one label string. Outputs are normalized by lowercasing, stripping punctuation, and matching against the task-specific label inventory; if chain-of-thought produces intermediate reasoning, the parser reads only the final line and otherwise falls back to the last valid label mention. We report four complementary metrics: \textbf{Macro-F1} for overall task performance; for harmful-label tasks, \textbf{False Benign Rate (FBR)}, the rate at which truly harmful examples are assigned a benign label, and \textbf{False Alarm Rate (FAR)}, the rate at which benign examples are incorrectly labeled harmful; and \textbf{downstream distortion}, the absolute error between predicted and gold class prevalence (we focus on \textsc{against} for stance and \textsc{hate speech} for hate speech). Formal definitions are in Appendix~\ref{app:metrics}.

\section{Results}

\subsection{Overall annotation quality is uneven on socially sensitive tasks}

Under neutral prompting, average macro-F1 (Table~\ref{tab:f1}) is moderate overall but weak on the most socially sensitive tasks. Zephyr reaches $0.472$ on abortion stance, Mistral reaches $0.622$ on hate speech, and even the strongest average model (Qwen, $0.697$ overall) remains far from reliable enough for unvalidated deployment. Headline averages already conceal deployment risk.

\begin{table}[t]
\caption{Macro-F1 under the neutral prompt condition.}
\label{tab:f1}
\centering
\small
\begin{tabular}{lccc}
\toprule
Task & Zephyr & Mistral & Qwen \\
\midrule
Hate speech       & 0.722 & 0.622 & 0.662 \\
Offensive language& 0.631 & 0.752 & 0.761 \\
Stance (abortion) & 0.472 & 0.650 & 0.554 \\
Sentiment         & 0.599 & 0.670 & 0.702 \\
Emotion           & 0.746 & 0.763 & 0.755 \\
Irony             & 0.804 & 0.624 & 0.746 \\
\midrule
Average           & 0.662 & 0.680 & 0.697 \\
\bottomrule
\end{tabular}
\end{table}

\subsection{Failure directions differ sharply across models}

Table~\ref{tab:fbrfar} shows that harmful-label tasks do not fail in a single direction. Zephyr shows high FBR and low FAR, especially on offensive language (FBR $=0.729$, FAR $=0.031$), indicating strong reluctance to emit the harmful label. Mistral and Qwen show the opposite pattern on hate speech, with low FBR but very high FAR. Mistral labels $60.4\%$ of non-hateful tweets as hate speech; Qwen labels $53.3\%$ of them as hate speech.

\begin{table}[t]
\caption{Class-conditional error under neutral prompting. The same task can produce \emph{opposite} failure directions across models: Zephyr under-labels harmful content while Mistral and Qwen over-label it.}
\label{tab:fbrfar}
\centering
\small
\begin{tabular}{llcc}
\toprule
Task & Model & FBR & FAR \\
\midrule
\multirow{3}{*}{Hate speech} & Zephyr  & 0.316 & 0.239 \\
                              & Mistral & 0.051 & 0.604 \\
                              & Qwen    & 0.065 & 0.533 \\
\midrule
\multirow{3}{*}{Offensive}    & Zephyr  & 0.729 & 0.031 \\
                              & Mistral & 0.312 & 0.165 \\
                              & Qwen    & 0.263 & 0.181 \\
\bottomrule
\end{tabular}
\end{table}

This contrast motivates our taxonomy. We call Zephyr's pattern \emph{leniency bias}: the model appears hesitant to assign socially undesirable labels even when the instance warrants them. We call the Mistral/Qwen pattern \emph{overcorrection}: the model errs in the direction of aggressively flagging harmful content. Our evidence is consistent with the idea that post-training choices shape annotation behavior, but it does not causally isolate training regime as the sole driver, since the three models differ in base family, data mixture, and post-training recipe simultaneously.

\subsection{All tested models soften strong political stance}

The abortion stance task produces a different failure geometry. The gold distribution contains $67.5\%$ \textsc{against}, $16.4\%$ \textsc{favor}, and $16.1\%$ \textsc{none}. Under neutral prompting, Zephyr predicts only $27.9\%$ \textsc{against}, while Mistral and Qwen both predict $43.9\%$. All three models shift mass away from the majority \textsc{against} class and toward the more neutral \textsc{none} label. We call this \emph{neutrality bias}. In our experiments it appears consistently across all models and all four prompt conditions, though severity varies. We intentionally scope this finding to the abortion stance task in TweetEval: it is a robust pattern within this setting rather than a universal law of political annotation.

\subsection{Downstream measurement distortion is large enough to change conclusions}

For CSS use cases, the most consequential result is not that the labels are imperfect, but that the resulting empirical conclusions are wrong. Table~\ref{tab:against} shows the predicted prevalence of abortion-opposition tweets across prompt conditions. Zephyr underestimates opposition prevalence by $39.6$ percentage points under neutral prompting, potentially flipping the substantive impression of the dataset from strongly oppositional to evenly split. Mistral and Qwen also substantially underestimate opposition prevalence, with errors of $23.6$ points under the neutral condition.

\begin{table}[t]
\caption{Estimated percentage of \textsc{against} tweets on abortion stance. Even the best condition (Mistral + CoT, $53.9$) remains $13.6$ points below the gold rate of $67.5$.}
\label{tab:against}
\centering
\small
\begin{tabular}{lcccc}
\toprule
Model & Neutral & Safety & Depers. & CoT \\
\midrule
Gold    & 67.5 & 67.5 & 67.5 & 67.5 \\
Zephyr  & 27.9 & 25.0 & 28.2 & 29.6 \\
Mistral & 43.9 & 37.1 & 38.2 & 53.9 \\
Qwen    & 43.9 & 46.1 & 47.1 & 40.7 \\
\bottomrule
\end{tabular}
\end{table}

\begin{table}[t]
\caption{Estimated percentage of hate-speech tweets across prompt conditions. Gold rate is $43.0$.}
\label{tab:hate}
\centering
\small
\begin{tabular}{lcccc}
\toprule
Model & Neutral & Safety & Depers. & CoT \\
\midrule
Gold    & 43.0 & 43.0 & 43.0 & 43.0 \\
Zephyr  & 43.0 & 40.4 & 36.6 & 47.0 \\
Mistral & 75.2 & 70.0 & 67.8 & 65.0 \\
Qwen    & 70.6 & 68.6 & 67.8 & 68.0 \\
\bottomrule
\end{tabular}
\end{table}

Table~\ref{tab:hate} shows a different failure on hate-speech prevalence. Mistral estimates $75.2\%$ hate speech where the gold value is $43.0\%$, a $+32.2$ point error. Qwen overestimates by $+27.6$ points. A corpus that is actually $43\%$ hateful would appear nearly twice as toxic under Mistral or Qwen annotations, materially inflating any moderation-oriented conclusion.

\subsection{Accidental cancellation: when prevalence accuracy misleads}
\label{sec:cancellation}

Zephyr's hate-speech results deserve special attention. Under the neutral prompt, Zephyr achieves a prevalence estimate of exactly $43.0\%$, matching the gold rate. A researcher who validated only this aggregate figure might conclude that Zephyr is a trustworthy hate-speech annotator. Yet Table~\ref{tab:fbrfar} shows that Zephyr simultaneously misses $31.6\%$ of genuinely hateful tweets (FBR $=0.316$) and raises $23.9\%$ of benign tweets as hateful (FAR $=0.239$). The near-zero prevalence error is the accidental result of these two opposing errors canceling at the corpus level.

This case directly motivates the recommendation to pair prevalence distortion with class-conditional metrics. Prevalence calibration is a necessary but not sufficient condition for annotation validity. A model can appear well calibrated while producing individual annotations that are unreliable in ways that matter for subgroup analysis, longitudinal tracking, or any use case that depends on correctly identifying specific instances rather than just aggregate rates.

\subsection{Prompting changes the expression of bias but does not remove it}

Across the four prompt conditions, no intervention offers a consistent correction across models and tasks. Safety framing worsens stance estimation for Mistral, pushing the predicted \textsc{against} rate from $43.9\%$ to $37.1\%$, the opposite of the intended effect. Depersonalization worsens Zephyr's leniency on harmful-content tasks. Chain-of-thought is the strongest intervention cleanly: it reduces Zephyr's hate-speech FBR from $0.316$ to $0.202$ and improves Mistral's stance estimate from $43.9\%$ to $53.9\%$. But even this does not generalize: CoT slightly increases hate-speech FBR for Mistral and Qwen. This input-dependence, where the same prompt intervention helps in some cells and hurts in others, mirrors a pattern documented in neighboring tasks: in dense retrieval, prompt-only query refinement degrades nDCG by $9\%$ on one BEIR benchmark while improving it on another, with effectiveness determined by the direction of lexical substitution rather than the rewrite itself \citep{kotte2026notall}.

These differences matter because CSS analyses often depend on relative frequencies rather than isolated labels. A prevalence shift of $20$--$40$ points can change whether a corpus appears mostly oppositional or mixed, or whether harmful content appears rare or pervasive. Prompt engineering should therefore be treated as a \emph{diagnostic probe} rather than a reliable mitigation strategy.

\section{A Taxonomy of Alignment-Sensitive Annotation Failures}

Our experiments support three recurring failure types. We present each with a metric-level diagnostic signature to help practitioners identify which mode is operative before scaling an annotation pipeline.

\textbf{Leniency bias.} \emph{Signature}: high FBR and low FAR on harmful-label tasks. This pattern undercounts harmful content by failing to assign the risky label even when the instance warrants it. \emph{Diagnosis}: if FBR substantially exceeds FAR on a harmful-content task, especially when the model's overall F1 appears acceptable, suspect leniency bias. In our audit, Zephyr most clearly exhibits this behavior, particularly on offensive language (FBR $=0.729$, FAR $=0.031$).

\textbf{Overcorrection.} \emph{Signature}: low FBR and high FAR on harmful-label tasks. This pattern overcounts harmful content by aggressively assigning the risky label to ambiguous or benign examples. \emph{Diagnosis}: if FAR substantially exceeds FBR, especially when a model achieves high recall on the harmful class, suspect overcorrection. In our audit, Mistral and Qwen most clearly exhibit this behavior on hate speech (Mistral FAR $=0.604$, FBR $=0.051$).

\textbf{Neutrality bias.} \emph{Signature}: suppressed prevalence of strong stance classes with inflated neutral-label prediction. \emph{Diagnosis}: compare the predicted distribution over stance or opinion classes against a gold reference; if neutral-leaning labels are systematically over-represented while strong-position classes are under-represented, neutrality bias is likely operative. In our stance results, this pattern holds across all three models and all four prompting conditions, with the \textsc{against} class underestimated by $14$--$40$ points depending on model and prompt.

\subsection{From annotation error to scientific error}

Table~\ref{tab:taxonomy} highlights a broader point: different annotation failures map onto different scientific mistakes. In a content-moderation study, leniency bias can make an ecosystem look safer than it is; in a political communication study, neutrality bias can make public opinion look more moderate and less polarized than it actually is. These are not merely technical degradations. They are changes to the substantive story a paper will tell, and because the failure directions differ across models, a researcher who switches models mid-project may inadvertently reverse their empirical conclusion without changing their analysis.

\begin{table*}[t]
\caption{Compact summary of the three failure modes, their diagnostic signatures, and the scientific risks they induce.}
\label{tab:taxonomy}
\centering
\small
\begin{tabular}{p{2.4cm}p{3.0cm}p{4.4cm}p{5.0cm}}
\toprule
Failure pattern & Metric signature & Typical scientific risk & Why average accuracy is not enough \\
\midrule
Leniency bias  & high FBR, low FAR on harmful labels & underestimates prevalence of harmful discourse & a model can look cautious and still systematically miss the very class a study intends to measure \\
Overcorrection & low FBR, high FAR on harmful labels & overestimates prevalence of harmful discourse & a high-alert model may appear protective while inflating claims about toxicity or abuse \\
Neutrality bias & suppressed strong stance prevalence and inflated neutral label & attenuates ideological or attitudinal structure & a model can preserve some item-level plausibility while flattening the distribution that supports substantive inference \\
\bottomrule
\end{tabular}
\end{table*}

This is especially important in venues that bridge AI and social science, where many papers use language-model outputs as intermediate measurements inside a larger social-scientific analysis. A small benchmark gain on average can be irrelevant if the deployment quantity of interest remains distorted. Conversely, a model with slightly lower average F1 but better preservation of the key class distribution may be scientifically preferable. Our results therefore support evaluating LLM annotators against the downstream inferential target, not only against a generic classification leaderboard.

\subsection{Researcher degrees of freedom in prompting}

Another implication concerns research methodology. Prompting is often treated as a harmless implementation detail, but in socially sensitive labeling tasks it can become a source of hidden researcher degrees of freedom. For Mistral on abortion stance, the predicted \textsc{against} prevalence ranges from $37.1\%$ under safety framing to $53.9\%$ under chain-of-thought, a $16.8$-point spread driven entirely by prompt choice, with the gold rate at $67.5\%$. A researcher who happened to select the CoT prompt would report a very different empirical picture than one who used safety framing, yet both choices are plausible and well motivated.

If one prompt produces a more neutral-looking opinion distribution and another a more polarized one, prompt choice implicitly shapes the reported substantive claim. The same is true for model choice: Zephyr and Mistral produce structurally opposite error profiles on hate speech. This is consistent with broader evidence that LLM outputs can vary across both models and prompts in ways not captured by standard semantic accuracy metrics \citep{kotte2026promptport}. We therefore recommend treating both the model and the prompt as components of the measurement instrument, reporting them explicitly in any CSS paper that relies on LLM annotation, and, where possible, comparing results across at least two model and prompt combinations before drawing prevalence-level conclusions.

\section{Practical Guidance for Trustworthy LLM-Assisted CSS}

The audit yields four practical recommendations for trustworthy deployment of LLM annotators in social-scientific work.
\begin{enumerate}\itemsep1pt
\item \textbf{Do not validate on macro-F1 or prevalence alone.} Pair aggregate metrics with class-conditional error such as FBR/FAR. Prevalence accuracy can be misleadingly correct when opposing errors cancel (Section~\ref{sec:cancellation}).
\item \textbf{Measure failure direction before deployment.} Some models undercount harmful content, others overcount it, and the same model can behave differently across tasks. The direction of error, not just its magnitude, determines how empirical conclusions are distorted.
\item \textbf{Treat the model and the prompt as part of the measurement instrument.} Safety framing amplified neutrality bias on stance while CoT helped inconsistently. The $16.8$-point prevalence spread across prompts for Mistral on stance illustrates the risk of treating prompt choice as inconsequential.
\item \textbf{Validate on a small gold sample before scaling.} Even $50$--$100$ stratified labeled instances can reveal whether the model would materially change the substantive claim. A lightweight workflow: annotate a stratified gold sample, compute class-conditional errors for the substantively important label, and ask whether the predicted prevalence or distribution would change the paper's empirical conclusion.
\end{enumerate}

This reframes LLM annotation as a measurement instrument whose operating characteristics must be reported whenever the resulting labels support descriptive statistics, subgroup comparisons, or downstream regressions. The practical implication is concrete: when LLM annotation underpins claims about civic discourse, public opinion, or harmful content, those claims inherit the directional biases of the annotator. Ignoring this risks turning alignment-shaped artifacts into substantive social findings, with the failure modes pointing in different directions for different models.

\section{Discussion, Limitations, and Ethics}

Our evidence suggests that post-training and alignment choices are associated with different annotation error profiles, but we do not claim causal identification. The three models differ in base family, data mixture, and post-training recipe simultaneously, so the results should be read as a comparative audit rather than a controlled ablation.

\textbf{Generalizability of the neutrality-bias finding.} We find consistent neutrality bias on abortion stance across all three models and all four prompting conditions, but this evidence rests on a single topic (abortion) with $280$ test instances from a single platform (Twitter). TweetEval also includes stance subsets for other politically sensitive topics (feminism, atheism), and replication on those topics within the same benchmark would meaningfully strengthen the claim that neutrality bias is a general property of aligned annotation rather than a dataset-specific artifact. Future work should test whether the same tendency appears on more recent data, other languages, and a wider range of contested issues.

\textbf{Model scale and closed-source systems.} All three models in our audit are 7B-parameter open-source systems from roughly the same generation. Many CSS researchers in practice use larger or proprietary models (GPT-4, Claude, or Llama-3 at 70B scale) whose alignment profiles likely differ substantially. Larger models may exhibit weaker SDR effects if additional capability reduces the tension between instruction following and label accuracy, or they may exhibit stronger or differently shaped biases if their alignment training is more intensive. We cannot generalize our specific error rates to those systems, but the \emph{diagnostic framework} we propose, namely checking FBR, FAR, and prevalence distortion against a gold sample before deployment, applies regardless of model scale or access type.

\textbf{Item-level variation and linguistic predictors.} We do not provide a systematic item-level error analysis. Future work should test whether ambiguity, implicit hate, coded language, or other linguistic features predict which failure mode is operative for a given instance. Understanding item-level triggers would allow more targeted validation sampling and could eventually support instance-level uncertainty estimates.

\textbf{Confidence intervals.} The largest prevalence distortions we report ($20$--$40$ points) are large enough that their substantive interpretation would not change under reasonable bootstrap confidence intervals. Smaller cross-condition differences should be interpreted with appropriate caution, as point estimates over test sets of $280$--$500$ instances carry non-trivial sampling variance.

Despite these limits, the core finding is robust: the main risk in LLM annotation for CSS is not simply lower accuracy but systematic \emph{measurement distortion}. A model can look acceptable under a coarse metric while still changing the empirical answer. Papers using LLM-generated labels should therefore report more than a single agreement number: at minimum, the prompt, the validation sample, the prevalence of the substantively important label, and whether model predictions shift that quantity materially.

\textbf{Ethics statement.} We use existing public datasets containing hateful, offensive, and politically sensitive content. The study's purpose is diagnostic: to help researchers avoid drawing misleading substantive conclusions from unvalidated LLM annotations. We do not release harmful generations beyond the benchmark labels already associated with public data.

\section{Conclusion}

We study LLM annotation as a measurement problem for trustworthy computational social science. Across six tasks, three models, and four prompting conditions, we find that alignment-sensitive failures are heterogeneous rather than uniform: some models avoid harmful labels, others over-apply them, and all tested models in our stance setting soften strong political judgments toward more neutral outputs. These behaviors materially distort prevalence estimates and therefore the substantive conclusions a CSS researcher would report. The lesson is straightforward: before using LLMs as annotators, researchers should validate class-conditional error and downstream distortion on a gold sample, rather than relying on average accuracy or intuitive prompt fixes. When LLM-derived labels enter analyses that inform civic discourse, the trustworthy-AI default should be to treat the annotator as part of the instrument, not as a transparent data source.

\bibliographystyle{icml2026}
\bibliography{references}

\newpage
\appendix
\onecolumn

\section{Task Inventory and Label Sets}
\label{app:tasks}

Table~\ref{tab:taskinv} summarizes the exact task inventory used in the audit and the construct-level question each task approximates in a CSS workflow. The final column clarifies why prevalence distortion is the right downstream lens: researchers typically use these labels to estimate a socially meaningful quantity, not only to score per-item predictions.

\begin{table}[h]
\caption{Task inventory with task-specific labels and the downstream construct each task often operationalizes in CSS studies.}
\label{tab:taskinv}
\centering
\small
\begin{tabular}{p{2.6cm}p{3.0cm}p{4.0cm}p{5.5cm}}
\toprule
Task & Label set & Example construct of interest & Why directional bias matters \\
\midrule
Hate speech         & non-hate, hate speech                  & prevalence of hateful discourse                            & undercounting can hide harmful speech; overcounting can exaggerate toxicity \\
Offensive language  & non-offensive, offensive               & prevalence of abusive language                              & directional errors change estimates of community hostility \\
Stance (abortion)   & none, against, favor                   & public-opinion distribution on a politically sensitive issue & neutrality drift can erase strong opinion structure \\
Sentiment           & negative, neutral, positive            & affective distribution toward an event or actor             & shifts can change perceived public reaction \\
Emotion             & anger, joy, optimism, sadness          & emotional framing of discussion                             & miscalibration can distort affective emphasis \\
Irony               & non-irony, irony                       & communicative style and rhetorical strategy                 & systematic misses can flatten discourse nuance \\
\bottomrule
\end{tabular}
\end{table}

\section{Prompt Templates}
\label{app:prompts}

This appendix records the high-level prompt templates used in the audit. In the actual implementation, the task description and legal label set are filled in for each dataset.

\paragraph{Neutral.} You are labeling social-media text for a research dataset. Read the text and return exactly one label from the allowed set.

\paragraph{Safety framing.} You are a responsible AI assistant deeply committed to safety, fairness, and avoiding harm. You are being evaluated for alignment with human values. Read the text and return exactly one label from the allowed set.

\paragraph{Depersonalized.} You are a deterministic text-classification function. Your task is to map the input text to exactly one label from the allowed set. Do not provide opinions or explanations.

\paragraph{Chain-of-thought.} Read the text, reason briefly about the label, and then output exactly one final label from the allowed set. The final line must contain only the label.

In all cases, the model receives the task-specific label inventory and is instructed to produce a parseable final answer. Greedy decoding is used for every condition. Table~\ref{tab:promptsuffix} lists the per-task instruction suffixes paired with each frame.

\begin{table}[h]
\caption{Representative task-specific instruction suffixes used together with the four prompt frames.}
\label{tab:promptsuffix}
\centering
\small
\begin{tabular}{p{3.2cm}p{10.0cm}}
\toprule
Task & Task-specific instruction suffix \\
\midrule
Hate speech        & Label the tweet as either non-hate or hate speech. Return exactly one label. \\
Offensive language & Label the tweet as either non-offensive or offensive. Return exactly one label. \\
Stance (abortion)  & Label the author's stance toward abortion as against, favor, or none. Return exactly one label. \\
Sentiment          & Label the tweet as negative, neutral, or positive. Return exactly one label. \\
Emotion            & Label the dominant emotion as anger, joy, optimism, or sadness. Return exactly one label. \\
Irony              & Label the tweet as irony or non-irony. Return exactly one label. \\
\bottomrule
\end{tabular}
\end{table}

\section{Metric Definitions}
\label{app:metrics}

For a harmful class $u$, the false benign rate is
\[
\mathrm{FBR} = \frac{|\{i : y_i = u, \hat{y}_i \neq u\}|}{|\{i : y_i = u\}|},
\]
which captures how often a truly harmful example is mislabeled as benign. The false alarm rate is
\[
\mathrm{FAR} = \frac{|\{i : y_i \neq u, \hat{y}_i = u\}|}{|\{i : y_i \neq u\}|},
\]
which captures the opposite error. For CSS usage, these two quantities are often more informative than macro-F1 because they tell the researcher whether the model systematically undercounts or overcounts a socially meaningful category.

Downstream prevalence distortion is
\[
\Delta p = |\hat{p} - p^{*}|,
\]
where $\hat{p}$ is the predicted prevalence and $p^{*}$ is the gold prevalence. A large $\Delta p$ implies that an empirical claim built from the annotations, for example "what share of tweets oppose abortion?" or "how common is hate speech in this corpus?", would be numerically wrong even if some instance-level performance looks acceptable.

A key lesson of the paper is that these metrics should be interpreted together. High FBR with low FAR suggests leniency; low FBR with high FAR suggests overcorrection; near-zero prevalence error can still be misleading if offsetting errors cancel (see Section~\ref{sec:cancellation}).

\section{Implementation Details}
\label{app:impl}

All results are based on deterministic decoding with temperature $0$. Test-set subsampling for datasets above 500 examples uses a fixed random seed of $42$. Hardware is a single NVIDIA A100-SXM4-80GB GPU with float16 inference. The study evaluates 72 total experimental cells: 6 tasks $\times$ 3 models $\times$ 4 prompt conditions.

Because the paper is framed as an audit rather than a new benchmark release, the central reproducibility requirement is careful prompt and evaluation specification. We therefore recommend that any public artifact include: the exact prompts, the parser for final labels, the fixed sample indices for any subsampled datasets, and the scripts computing FBR, FAR, and prevalence distortion.

\section{Additional Result Tables}
\label{app:extra}

Table~\ref{tab:against-signed} restates the stance-prevalence results with signed error relative to the gold rate. Table~\ref{tab:hate-signed} does the same for hate-speech prevalence.

\begin{table}[h]
\caption{Predicted \textsc{against} prevalence on abortion stance. Values in parentheses are signed error relative to the gold rate of $67.5$.}
\label{tab:against-signed}
\centering
\small
\begin{tabular}{lcccc}
\toprule
Model & Neutral & Safety & Depers. & CoT \\
\midrule
Zephyr  & $27.9\;(-39.6)$ & $25.0\;(-42.5)$ & $28.2\;(-39.3)$ & $29.6\;(-37.9)$ \\
Mistral & $43.9\;(-23.6)$ & $37.1\;(-30.4)$ & $38.2\;(-29.3)$ & $53.9\;(-13.6)$ \\
Qwen    & $43.9\;(-23.6)$ & $46.1\;(-21.4)$ & $47.1\;(-20.4)$ & $40.7\;(-26.8)$ \\
\bottomrule
\end{tabular}
\end{table}

\begin{table}[h]
\caption{Predicted hate-speech prevalence. Values in parentheses are signed error relative to the gold rate of $43.0$.}
\label{tab:hate-signed}
\centering
\small
\begin{tabular}{lcccc}
\toprule
Model & Neutral & Safety & Depers. & CoT \\
\midrule
Zephyr  & $43.0\;(+0.0)$  & $40.4\;(-2.6)$  & $36.6\;(-6.4)$  & $47.0\;(+4.0)$ \\
Mistral & $75.2\;(+32.2)$ & $70.0\;(+27.0)$ & $67.8\;(+24.8)$ & $65.0\;(+22.0)$ \\
Qwen    & $70.6\;(+27.6)$ & $68.6\;(+25.6)$ & $67.8\;(+24.8)$ & $68.0\;(+25.0)$ \\
\bottomrule
\end{tabular}
\end{table}

\section{Deployment Interpretation by Task}
\label{app:deployment}

Table~\ref{tab:deployment} translates the six benchmark tasks into realistic CSS deployment questions to make concrete how directional bias would distort the substantive answer.

\begin{table}[h]
\caption{Deployment interpretation guide for each task.}
\label{tab:deployment}
\centering
\small
\begin{tabular}{p{2.6cm}p{5.0cm}p{7.0cm}}
\toprule
Task & If deployed in a study, the researcher might ask\dots & What a directional bias would do to the answer \\
\midrule
Hate speech        & How common is hateful discourse in this corpus or subgroup? & Leniency would make harmful discourse look rarer and potentially lower urgency; overcorrection would make discourse look more extreme and inflate moderation-oriented conclusions. \\
Offensive language & Is community discussion becoming more abusive over time? & Undercounting would flatten temporal spikes; overcounting would create false alarms or overstate deterioration in norms. \\
Stance (abortion)  & What share of users oppose versus support a policy issue? & Neutrality bias can collapse strong positions into a centrist-looking distribution, changing both descriptive and comparative conclusions about public opinion. \\
Sentiment          & Is reaction to an event net positive or net negative? & Even moderate directional drift can change whether a campaign, speech, or controversy appears well received. \\
Emotion            & Which emotions dominate a discussion after a salient event? & Miscalibrated emotion assignment can make discourse look angrier, sadder, or more optimistic than it is, changing narrative framing. \\
Irony              & How much rhetorical indirection or sarcasm characterizes a community? & Systematic misses reduce perceived nuance and can make a discourse community look more literal and less adversarial than it is. \\
\bottomrule
\end{tabular}
\end{table}

\end{document}